\documentclass[10pt,aps,twocolumn,prd,superscriptaddress,showpacs,nofootinbib,noshowkeys,floatfix,preprintnumbers]{revtex4-1}

\usepackage{graphics,graphicx}
\usepackage{amsmath, amssymb}
\usepackage{multirow}
\usepackage{color}
\usepackage{hyperref}
\usepackage[normalem]{ulem}  

\usepackage{ulem}
\usepackage{color}

\renewcommand\sout{\bgroup \color{blue} \ULdepth=-.5ex \ULset}

\newcommand*\dif{\mathop{}\!\mathrm{d}}

\begin{document}

\title{Repulsive interactions and their effects on the thermodynamics of hadron gas
}

\date{\today}
\author{Pok Man Lo}
\affiliation{Institute of Theoretical Physics, University of Wroclaw,
PL-50204 Wroc\l aw, Poland}
\affiliation{Extreme Matter Institute EMMI, GSI,
Planckstr. 1, D-64291 Darmstadt, Germany}
\author{Bengt Friman}
\affiliation{GSI, Helmholzzentrum f\"{u}r Schwerionenforschung,
Planckstr. 1, D-64291 Darmstadt, Germany}
\author{Micha\l {} Marczenko}
\affiliation{Institute of Theoretical Physics, University of Wroclaw,
PL-50204 Wroc\l aw, Poland}
\author{Krzysztof Redlich}
\affiliation{Institute of Theoretical Physics, University of Wroclaw,
PL-50204 Wroc\l aw, Poland}
\affiliation{Extreme Matter Institute EMMI, GSI,
Planckstr. 1, D-64291 Darmstadt, Germany}
\author{Chihiro Sasaki}
\affiliation{Institute of Theoretical Physics, University of Wroclaw,
PL-50204 Wroc\l aw, Poland}

\begin{abstract}
	We compare two approaches in modeling repulsive interactions among hadrons: the excluded volume approximation and the S-matrix formalism. These are applied to study the thermodynamics of the $\pi N \Delta$ system. It is shown that the introduction of an extraneous repulsion between pions and nucleons via the excluded volume approach, in addition to the interaction that generates the $\Delta$-resonance, is incompatible with the analysis based on the physical phase shift of pion-nucleon scattering in the $P_{33}$ channel. { This finding suggests that the repulsive force between hadrons is  interaction-channel dependent and is hence unlikely to be captured by a single phenomenological parameter. The S-matrix approach employed here can be used, however,  to provide useful estimates of the magnitude of the effective eigenvolume.}
\end{abstract}




\maketitle 

\section{modeling repulsive interactions among hadrons}

One consequence of confinement in QCD is that physical observables admit an interpretation in terms of hadronic states. This means that at low temperatures the theory can be written in terms of ground state hadrons and their resonances. Such an approach for describing the thermodynamics of QCD is adopted by the hadron resonance gas (HRG) model~\cite{BraunMunzinger:2003zd}, where all known hadrons are included. The model assumes that resonance formation governs the thermodynamics of the confined phase, and as a first approximation, treats the resonances as point-like, neglecting their mutual interactions.

However, a closer inspection of the individual interaction channels of hadrons reveals that not all resonances should be incorporated as zero-width particles in the HRG partition sum~\cite{Raju}. This is due to the fact that some of the attractive channels are counteracted by a corresponding repulsive channel with a different isospin. This point is stressed in recent studies of $\pi \pi$ and $\pi K$ scattering~\cite{Broniowski,kappa}, where a strong cancellation between different isospin channels in the S-wave is demonstrated. Consequently, the contribution of the two lightest scalar resonances, i.e., the $f_0(500)$ and the (not yet confirmed) $K^{*}_0(800)$, to the HRG partition function requires additional care.

It is also necessary to account for the non-ideal behavior of the hadron gas. In particular, the fact that hadrons are not point-like particles but extended objects means that they cannot approach arbitrarily close to each other without interacting. Inspired by the short-range repulsion between nucleons, it is generally assumed that this interaction is repulsive~\footnote{We stress that this assumption is questionable. For instance, the nucleon-antinucleon interaction is probably attractive at short distances \cite{Andronic:2012ut}. Nevertheless, in this paper we restrict the discussion to the customarily assumed repulsive short-range interactions.}, which at a given temperature and chemical potential tends to lower the thermodynamic pressure. Such a repulsive interaction among the hadrons is commonly implemented by introducing an excluded volume, first suggested by Hagedorn, Rafelski and Gorenstein {\it et al.} \cite{Hagedorn:1980kb, Hagedorn:1982qh, Gorenstein:1981fa}, and later reformulated in a thermodynamically self-consistent version~\cite{Rischke, Kapusta, Albright:2015uua, Andronic:2012ut, Zalewski:2015yea}. These models introduce the particle eigenvolume as a model parameter. The eigenvolume is usually assumed to be a constant or inferred from the bag-model \cite{Kapusta}.

The S-matrix formalism, on the other hand, provides a consistent theoretical framework for implementing the mutual interaction among the hadrons~\cite{Dashen1, Weinhold:1997ig}. Within this approach, the interaction contribution to the thermodynamic pressure is, to leading order in the relativistic virial expansion, determined by the phase shifts for two-particle scattering. This scheme provides a natural framework to account for both attractive and repulsive hadronic forces. The phase shifts can be computed theoretically from hadron-hadron interactions or obtained by analyzing scattering experiments. The former offers a flexible theoretical tool for exploring the consequences of repulsive forces on the thermodynamics of a hadron gas.
	
In this paper, we compare two approaches to modeling repulsive interactions among hadrons: the excluded volume approximation and the S-matrix formalism. These are used to study the contribution of repulsive forces on the thermodynamics of the $\pi N \Delta$ system. The effect of an excluded volume on the equation of state of a hadronic system has been extensively studied at finite temperature and baryon density \cite{Rischke, Kapusta, Albright:2015uua, Andronic:2012ut, Zalewski:2015yea}. In this work we focus on the significant differences between the two approaches in computing the temperature dependence of thermal observables, in particular, the thermodynamic pressure and  the variance of net baryon number fluctuations.

We show that the introduction of an extraneous repulsive force in the $P_{33}$ channel of $\pi  N$ scattering leads to a distortion of the phase shift of the channel. Therefore, the undifferentiated application of excluded volume effects in all interaction channels is highly questionable. Moreover,  such a scheme leads to  a  suppression of the thermodynamic pressure and other derived observables, compared to a consistent implementation of interactions in the S-matrix approach based on phenomenological phase shifts.

The paper is organized as follows: In the next Section we introduce different  models of repulsive interactions. In Section III we discuss their  influence  on thermodynamics   in the $\pi N\Delta$ system.  In the last section we present our conclusions.


\section{Formulation of repulsive interactions}

\subsection{The excluded volume approximation}

	Based on phenomenological nucleon-nucleon interactions, it is commonly assumed that all hadrons repel each other at short distances. In the equation of state of hot and dense strongly interacting matter, this is often implemented by introducing an excluded volume~\cite{Rischke, Kapusta, Albright:2015uua, Andronic:2012ut, Zalewski:2015yea}. This leads to an equation of state similar to the van der Waals equation for non-ideal gas. Following Ref.~\cite{Rischke}, a thermodynamically self-consistent formulation of the model can be summarized as

	\begin{align}
	\label{eqn:ev}
	P^{\rm ev}(T,\mu) = P^{0}(T,\tilde{\mu} = \mu - v_0 \, P^{\rm ev}(T,\mu)),
	\end{align}
	
	\noindent where $v_0$ is the eigenvolume of the particle, $P^{0}$ is the pressure function of the original ideal system, and $P^{\rm ev}$ is the pressure function including the excluded volume effect.

	Note that $P^{0} $ is expected to be a known function of temperature and chemical potential, and Eq.~\eqref{eqn:ev} defines an implicit equation for $P^{\rm ev}(T,\mu)$. The latter has to be solved self-consistently, and from this other thermal observables can be derived.

	To demonstrate we consider the introduction of a constant excluded volume $v_0$ to an ideal gas of hadrons. In this case,

	\begin{align}
	\label{eqn:p_id1}
	P^{0}(T,\tilde{\mu}) &= \sum_\alpha P^{id}_\alpha (T, \tilde{\mu_\alpha}) \nonumber \\
	\tilde{\mu_\alpha} &= \sum_I \mu_I B^\alpha_I - v_0 P^{\rm ev}(T, \mu) \\
	I &= B, S, Q. \nonumber
	\end{align}

	\noindent Here $P^{id}_\alpha$ is the standard ideal gas expression of the thermodynamic pressure for species $\alpha$,

	\begin{align}
	\label{eqn:p_id2}
	P^{id}_\alpha(T,\tilde{\mu}) = \pm \int \frac{\dif^3 p}{(2 \pi)^3} \, \ln(1 \pm  e^{- (\sqrt{p^2 + m_\alpha^2}-\tilde{\mu})/T}),
	\end{align}

	\noindent where the $(\pm)$ sign applies to fermions and bosons respectively, $B^\alpha_I$'s are the conserved charges of particle $\alpha$, and $\mu_I$ being the chemical potentials associated with the conserved charges $I = B, S, Q$. In the particle sum the species index $\alpha$ distinguishes between a particle and an anti-particle.

\begin{figure*}[ht!]
	\centering
 \includegraphics[width=0.497\textwidth]{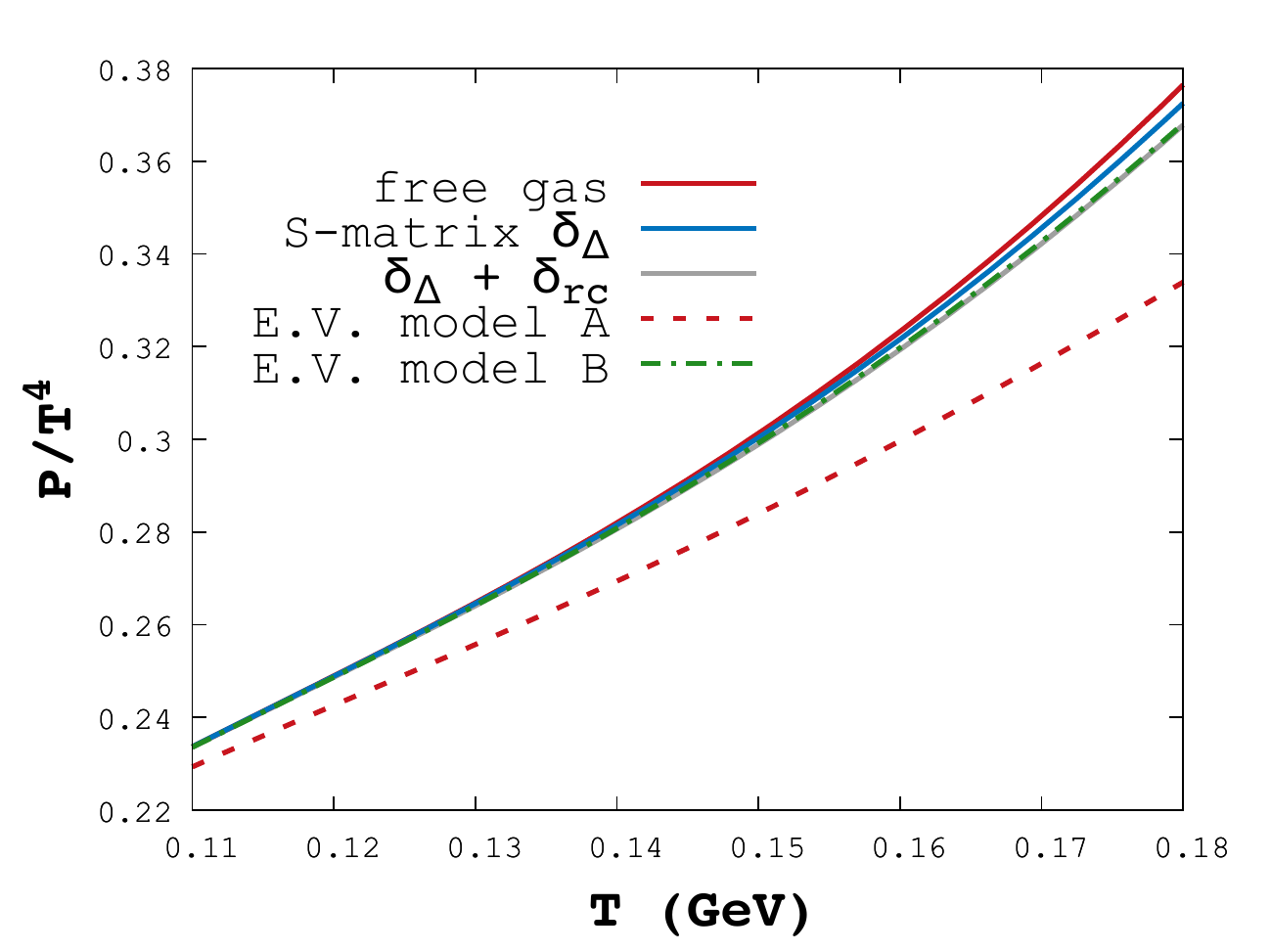}
 \includegraphics[width=0.497\textwidth]{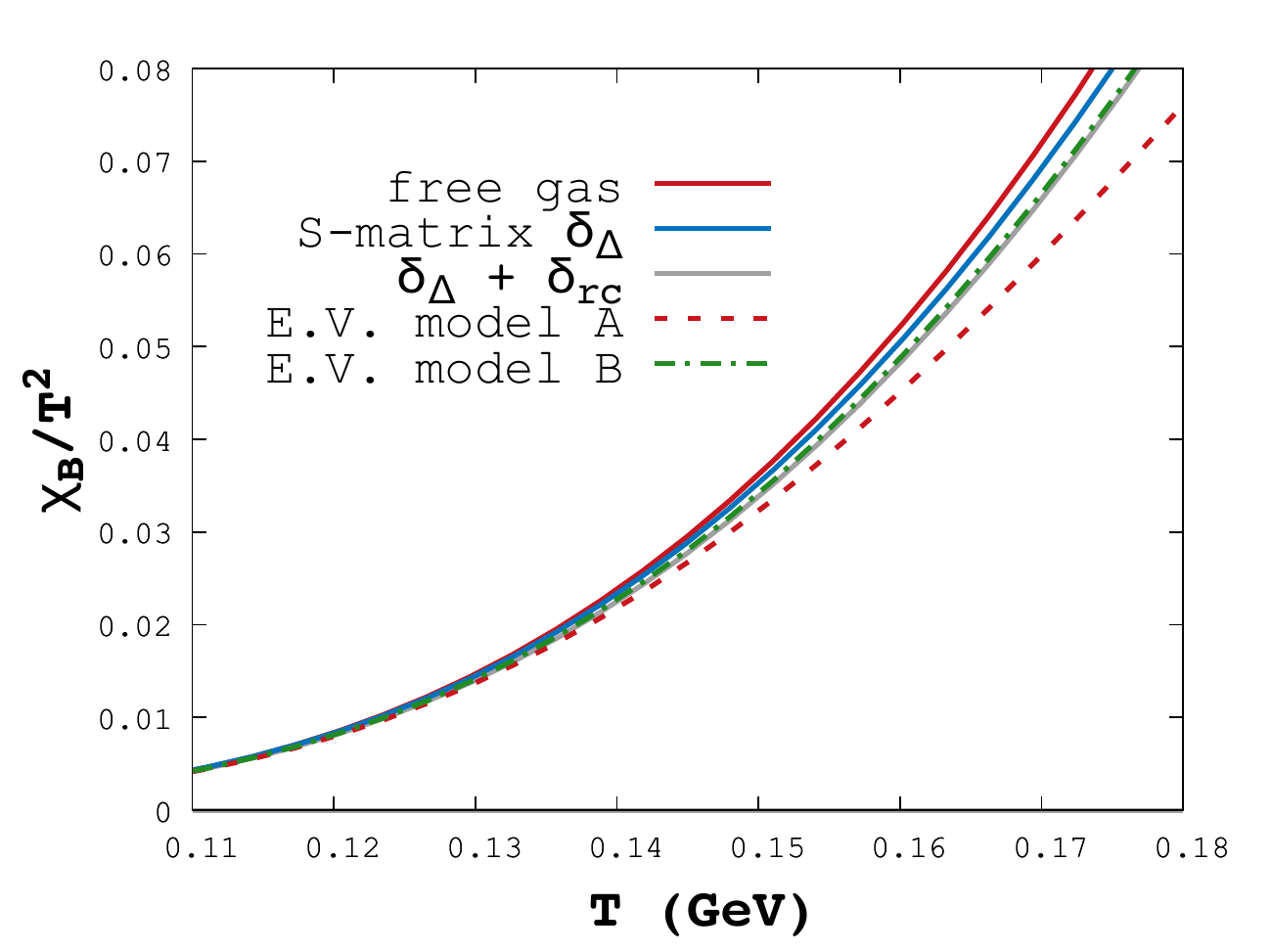}
	\caption{(color online). The pressure and baryon number susceptibility for the $\pi N \Delta$ system under various schemes, all scaled to dimensionless units, versus temperature at zero chemical potential. Free gas corresponds to an ideal gas of pions, nucleons and $\Delta$-baryons. S-matrix $\delta_\Delta$ makes use of the empirical $\pi N$($P_{33}$) phase shift data. The {\it scheme $\delta_\Delta + \delta_{rc}$} is a similar S-matrix treatment with an additional P-wave hard-core phase shift imposed. {\it Model A} of the excluded volume approach assigns a universal eigenvolume of radius $r_0 = 0.3$ fm to all hadrons, while {\it model B} is a similar implementation but imposing repulsion only between pions and nucleons. Details of the schemes are discussed in the text.}
 \label{fig:one}
\end{figure*}

	The full excluded-volume pressure $P^{\rm ev}$ for this system can be solved by applying $P^{0}$ from Eqs.~\eqref{eqn:p_id1}~-~\eqref{eqn:p_id2} to Eq.~\eqref{eqn:ev}. Derivation of other thermodynamical observables is straightforward \cite{Kapusta, Albright:2015uua,Andronic:2012ut}. Here we give the explicit expressions particularly for the net baryon number $n^{\rm ev}_B$ and the baryon number susceptibility $\chi^{\rm ev}_B$:

	\begin{align}
	\label{eqn:n_ev}
	\begin{aligned}
	n^{\rm ev}_B(T,\mu) &= \frac{\partial}{\partial \mu_B} P^{\rm ev}_B(T,\mu) \\
		       	    &=\frac{\sum_\alpha B^\alpha n^{\rm id}_\alpha}{1 + v_0 \sum_\alpha n^{\rm id}_\alpha}
	\end{aligned}
	\end{align}

	\begin{align}
	\label{eqn:chi_ev1}
	\begin{aligned}
	\chi^{\rm ev}_B(T,\mu) &= \frac{\partial}{\partial \mu_B} n^{\rm ev}_B(T,\mu) \\
							   &= \frac{ \sum_\alpha \chi^{\rm id}_\alpha B_\alpha^2 }{1 + v_0 \sum_\alpha n^{\rm id}_\alpha} + \chi_B^\prime
	\end{aligned}
	\end{align}

	\begin{align}
	\label{eqn:chi_ev2}
	\begin{aligned}
	\chi_B^\prime &=\frac{ -2 v_0 n^{\rm ev}_B(T,\mu) \sum_\alpha \chi^{\rm id}_\alpha B_\alpha + v_0^2 n^{\rm ev}_B(T,\mu)^2 \sum_\alpha \chi^{\rm id}_\alpha}{1 + v_0 \sum_\alpha n^{\rm id}_\alpha}
	\end{aligned}
   	\end{align}

	\noindent where $n^{\rm id}_\alpha$ and $\chi^{\rm id}_\alpha$ are given by the standard ideal gas expression for number density and particle-number susceptibility, respectively:

	\begin{align}
		n^{\rm id}_\alpha &= n^{\rm id}_\alpha(T, \tilde{\mu}) = \frac{\partial}{\partial \tilde{\mu}} P^{\rm id}(T, \tilde{\mu})
	\end{align}

	\noindent and

	\begin{align}
		\chi^{\rm id}_\alpha &= \chi^{\rm id}_\alpha(T, \tilde{\mu}) = \frac{\partial}{\partial \tilde{\mu}} n^{\rm id}(T, \tilde{\mu}).
	\end{align}

	\noindent Note that the various ideal gas quantities $\mathcal{O}^{\rm id}$ are evaluated at the shifted chemical potential $\tilde{\mu}$, defined in Eq.~\eqref{eqn:ev}. The net-baryon number vanishes as expected at $\mu = 0$ due to the cancellation between the contributions from particle and anti-particle. At the same limit, $\chi_B^\prime$ vanishes, and the baryon number susceptibility has a similar structure as the ideal gas counterpart, albeit the evaluation at the shifted chemical potential $\tilde{\mu}$ and the correction factor $(1 + v_0 \sum_\alpha n^{\rm id}_\alpha)$ from the excluded volume effect.

	\subsection{S-matrix approach for hard-sphere gas}

	The determination of the thermodynamic pressure within the S-matrix approach is a two-step process. Firstly, one must specify the phase shift for the interaction channel. This may be obtained empirically from the relevant scattering experiment or calculated from a phenomenological model. Secondly, one must compute the thermodynamic observables from the phase shift. This can be achieved by applying the relativistic virial expansion introduced by Dashen {\it et al.} \cite{Dashen1}.

	Here we demonstrate the method by studying a mixture of bosons and fermions with a hard-core type potential between the bosonic and fermionic constituents.
	The quantum mechanical problem of hard-sphere scatterings is rather standard, and we simply quote the result of the phase shift $\delta_l$ for arbitrary orbital angular momentum $l$ \cite{Shiff}:

	\begin{align}
	\label{eqn:ps}
	\delta_l = \arctan \left(\frac{j_l(q a_S)}{n_l(q a_S)}\right)
	\end{align}

	\noindent where $j_l$ and $n_l$ are respectively the first and the second kind of spherical Bessel function, $q$ is the relative momentum in the reduced system, and $a_S$ is the scattering length. For small $x = q a_S$, one gets
	
	\begin{align}
	\label{eqn:ps}
	\delta_l \approx \frac{-x^{2l +1} }{(2l+1)!((2l-1)!!)^2},
	\end{align}

	\noindent displaying the expected behavior of $\delta_l \propto (q a)^{2l +1} $ near the threshold.

	Having obtained an analytic expression for the phase shift of a hard-core potential, we proceed to compute its contribution to the thermodynamic pressure. The first correction to pressure due to interaction is given by the second virial coefficient, which is expressible in terms of the scattering phase shift via the famous formula of Beth and Uhlenbeck \cite{Beth:1937zz}. The interaction contribution to the pressure then reads

	\begin{figure*}[t]
 	\includegraphics[width=0.497\textwidth]{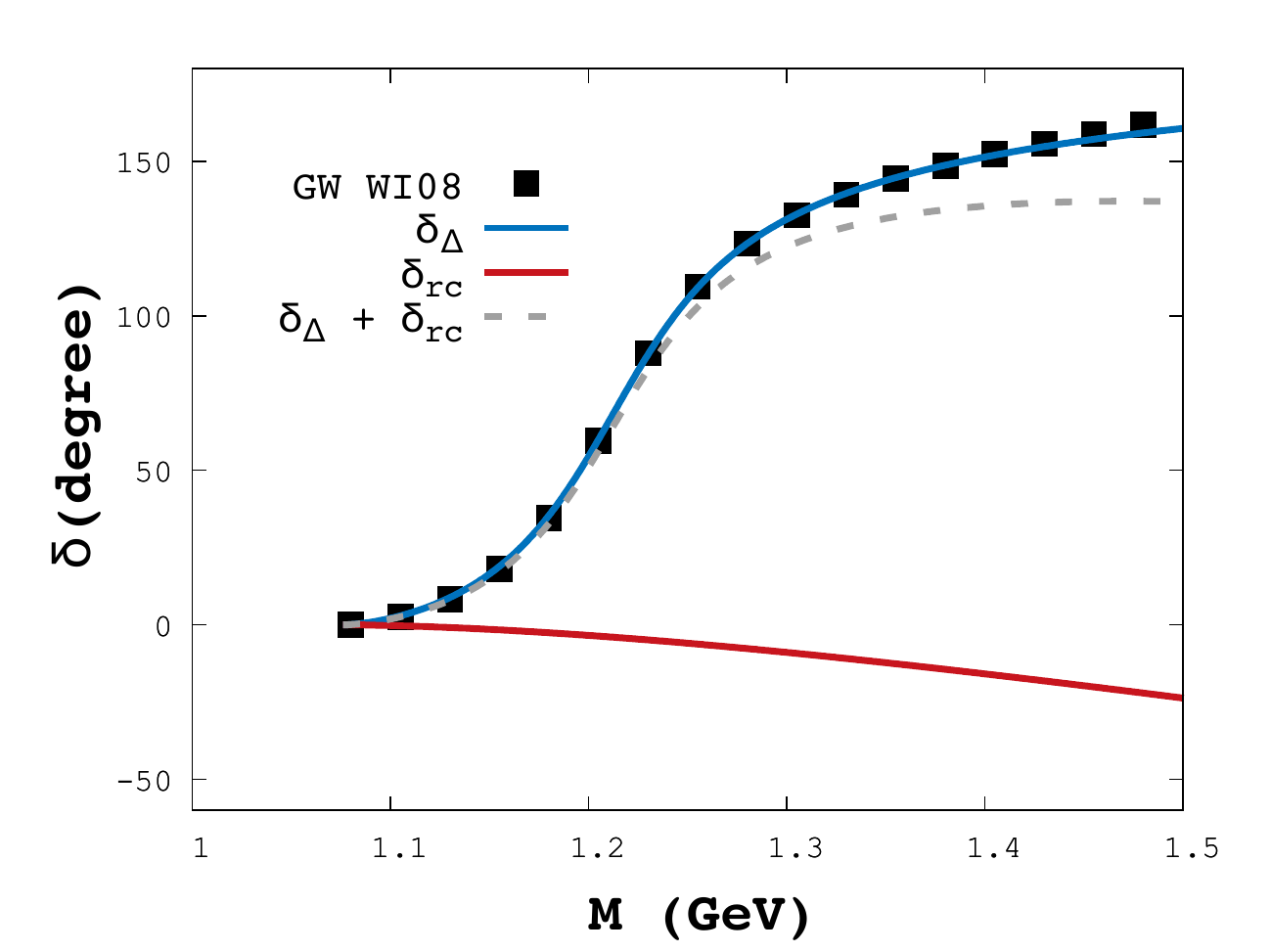}
 	\includegraphics[width=0.497\textwidth]{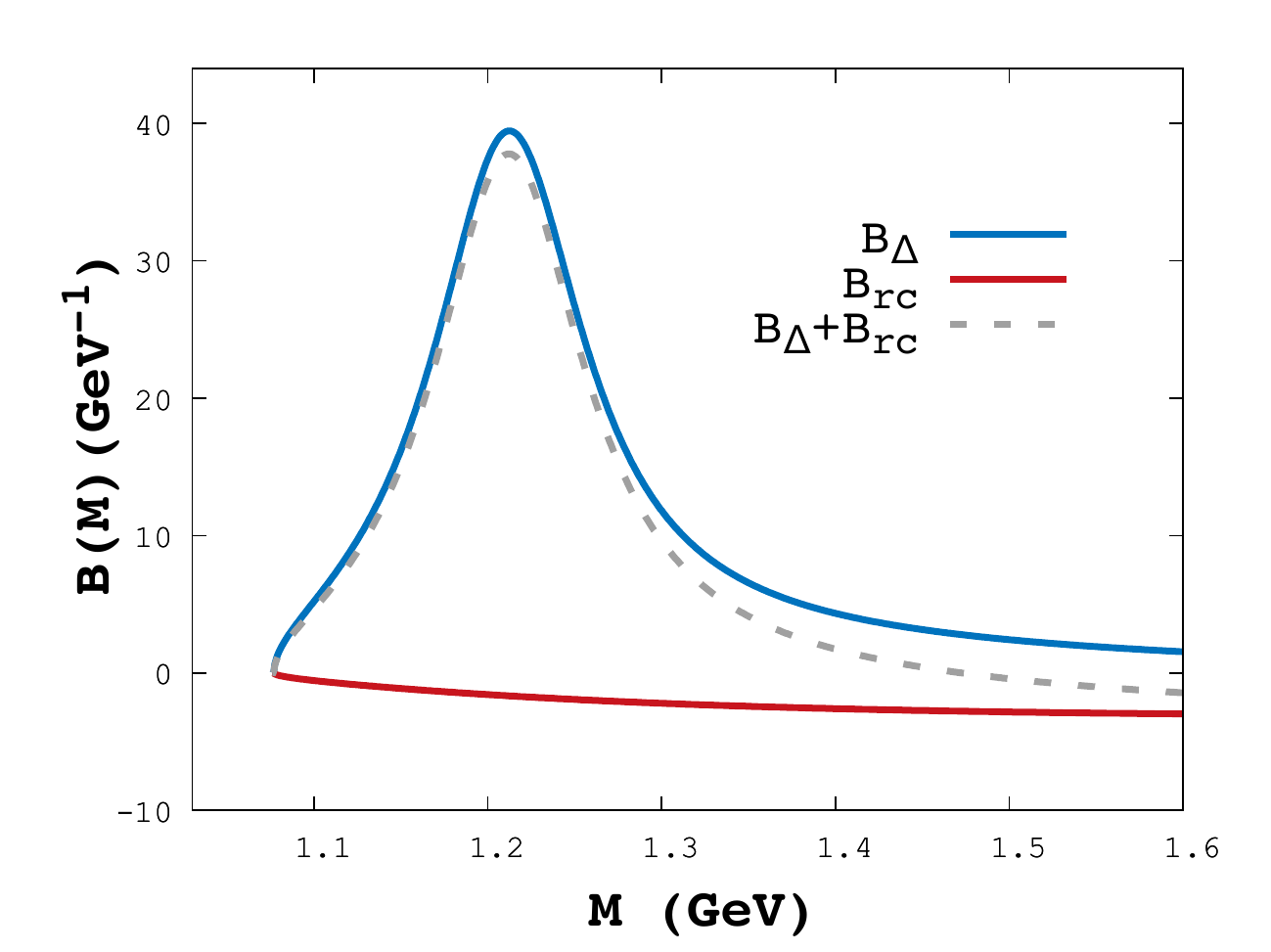}
	\caption{(color online). Left-hand figure: The $\pi N$ phase shifts in the $P_{33}$ channel,  as a function of center-of-mass energy $M$. The  $\delta_\Delta$ is from Eq. (14), and $\delta_{rc}$  from  Eq. (18). Right-hand figure:  The corresponding effective spectral functions $\mathcal{B}(M)$ calculated from Eq.~\eqref{eqn:eff_sf}, as a function of $M$. The symbols, GW WI08,  are  the empirical phase shift  values from Ref. \cite{Workman:2012hx}. }
 	\label{fig:two}
	\end{figure*}

	\begin{align}
	\begin{aligned}
	\label{eqn:p_bu}
	\Delta P^{\rm bu}_{\rm int.} &= d_g \int \frac{\dif M}{2 \pi} \, \mathcal{B}_l(M) \, \times  \\
			  & \int \frac{\dif^3 k}{(2 \pi)^3}  \, T \ln ( 1 + e^{-\beta (E(k; M)-\mu)} )
	\end{aligned}
	\end{align}

	\noindent where

	\begin{align}
	\begin{aligned}
	\label{eqn:eff_sf}
	\mathcal{B}_l(M) &= 2\,  \frac{\dif}{\dif M} \delta_l  \\
	E(k;M) &= \sqrt{k^2 + M^2} \\
	\mu &= \mu_1 + \mu_2.
	\end{aligned}
	\end{align}


	\noindent Here we have considered the case for the scattering between a fermion (with mass $m_1$ and chemical potential $\mu_1$) and a boson (with mass $m_2$ and chemical potential $\mu_2$). The angular momentum of this  two-body system is denoted by $l$.

	The interacting pressure in Eq.~\eqref{eqn:p_bu} clearly displays the statistical averaging (the integral over $k$) over an ensemble of effective particles of mass $M$  and the degeneracy factor  $d_g$. It also contains the dynamical structure (the integral over $M$) describing the correlation of the microscopic constituents. The Fermi statistics of the combined system, i.e. the use of $\ln ( 1 + e^{-\beta (E(k; M)-\mu)})$, is a consequence of the resummation of exchange diagrams from the higher virial terms \cite{Dashen1, Dashen2}. Finally, $\mathcal{B}_l(M)$ can be interpreted as an effective spectral function, which describes the change in the density of states due to interaction.

	\section{Application to $\pi N \Delta$ system}

	We now apply the two formulations of the repulsive force to study the thermodynamics of the $\pi N \Delta$ system. For this purpose, we consider the isospin $I = 3/2$, P-wave channel of the $\pi N$ scattering.

	To calculate the pressure within the excluded volume approximation, we identify the constant eigenvolume $v_0$ with~\cite{Andronic:2012ut}

	\begin{align}
		v_0 &= \frac{16}{3} \pi r_0^3,
	\end{align}

	\noindent and fix $r_0 = 0.3 \, {\rm fm}$. Once the excluded volume is specified, the pressure can be obtained by solving Eq.~\eqref{eqn:ev} self-consistently ({\it model A}).

	For an analogous study within the S-matrix approach, we make use of the empirical $P_{33}$ phase shift~\cite{Workman:2012hx} in the elastic region,  $M \lesssim 1.4 \, {\rm GeV}$. The phase shift can be expressed by the following phenomenological equation inspired by a one-loop perturbative calculation of $\Delta$ self-energy,
	
\begin{align}
\label{fit_formula}
\begin{aligned}
	\delta_{33}(M) &= \tan^{-1}\left(- \frac{2}{3 M} \, \frac{{q}^3}{M^2-m_0^2} \, \frac{\alpha_0}{1+ \Pi(q)} \right) \\
	\Pi(q) &= c_1 \, {q}^2 +c_2 \, {q}^4,
\end{aligned}
\end{align}

\noindent where $q$ is the momentum of the scattering constituents in the center-of-mass frame, expressible in $M$,  as

\begin{align}
	q(M) = \frac{M}{2} \sqrt{1-\frac{(m_N+m_\pi)^2}{M^2}} \sqrt{1-\frac{(m_N-m_\pi)^2}{M^2}},
\end{align}

\noindent and $\alpha_0 = 45.37  $, $m_0 = 1.2325 \, {\rm GeV}$, $c_1 = 16.7 \, {\rm GeV^{-2}}$ and $c_2 = 65.6 \, {\rm GeV^{-2}}$,  are model parameters. These are chosen to reproduce not only the phase shift data but also the P-wave scattering length,

\begin{align}
	{a^{\frac{3}{2}}}_{\frac{3}{2}} = \frac{\delta_{33}}{q^3} \bigg|_{q\rightarrow0} = 0.21 \, m_\pi^{-3},
 \end{align}

 \noindent  which is compatible with  the experimental value~\cite{Dumbrajs:1983jd},  $ 0.21(1) \, m_\pi^{-3} $.   This condition is essential for the correct description of the low invariant-mass behaviors of the density function $\mathcal{B}$ in Eq.~\eqref{eqn:eff_sf} .

In Fig.~\ref{fig:one}, we present the numerical results for temperature-scaled pressure $(P/T^4)$ and baryon number susceptibility,  $\chi_B/T^2$,  under different approximation schemes, at vanishing chemical potentials. From the left-hand figure, we observe that the S-matrix result on the thermodynamic pressure is comparable to that of a free gas of pions, nucleons and $\Delta$-baryons. The minute difference becomes more appreciable when one focuses only on the baryonic sector, e.g. studying the observable $\chi_B/T^2$, shown in Fig.~\ref{fig:one}-right.  This agreement, however, is only accidental, and should not be interpreted as a justification for the zero-width treatment of $\Delta$ baryons. In fact,  $\Delta(1232)$ is a broad resonance (see Fig.~\ref{fig:two}), and the criteria of effective elementarity suggested by Dashen {\it et al.} \cite{Dashen3} does not apply. As explained in Ref.~\cite{Weinhold:1997ig}, the effective spectral function $\mathcal{B}$ contains both the standard spectral function contribution ($\Delta$) and the contribution from a correlated $\pi N$ pair. The latter leads to an enhancement in the thermal observables above the corresponding Breit-Wigner limit. The combination of the two effects brings about the current condition of an apparent agreement with the free gas result.

The empirical phase shift suggests that the repulsive interaction within this channel, if exists, is rather weak. On the other hand, as seen in Fig.~\ref{fig:one},  the excluded volume approach yields a much stronger modification to the thermodynamic pressure. The corresponding difference in the predictions of baryon number susceptibility is not as prominent. In this case the correction from the excluded volume approach is only moderate, since its magnitude,  to leading order in $v_0$, is proportional to the ideal gas value of $\chi_B$, which is much smaller than the value of the pressure in the absence of repulsion.

The standard implementation of the excluded volume approach ({\it model A}) introduces repulsions among all hadron species. Consequently, in $\pi N\Delta$ system, in addition to the repulsion between pions and nucleons, the repulsion among pions are also included. For  further model comparison with the S-matrix approach, we follow Ref. \cite{Satarov:2016peb},  and introduce  excluded volume effects only between pions and nucleons ({\it model B}). In this case,

\begin{align}
	\begin{split}
	P^{ev} =& P_\pi + P_{N} + P_{\bar{N}} + P_\Delta + P_{\bar{\Delta}}  \\
	=&P^0_{\pi}(T, \tilde{\mu_\pi} = \mu_\pi - v_0 (P_N + P_{\bar{N}}))   + \\
	&P^0_{N}(T, \tilde{\mu}_N = \mu_N - v_0 P_\pi)  + \\
	&P^0_{\bar{N}}(T, \tilde{\mu}_{\bar{N}} = \mu_{\bar{N}} - v_0 P_\pi)  + \\
	&P^0_\Delta(T,\mu_\Delta) + P^0_{\bar{\Delta}}(T,\mu_{\bar{\Delta}})),
	\end{split}
        \label{eqn:modelB}
\end{align}

	\noindent and similar to Eq.~\eqref{eqn:p_id1}, the system of equations have to be solved self-consistently. The corresponding $\chi_B$ may be explicitly derived but the resulting expression is not particularly illuminating. Instead we resort to a direct numerical computation of $\mu_B$-derivatives of $P^{ev}$ in Eq.~\eqref{eqn:modelB}.

It is interesting to compare {\it model B} with a modified S-matrix treatment ({\it scheme $\delta_\Delta+\delta_{rc}$ }). Here we impose an additional repulsive force on the previous result in Eq.~\eqref{fit_formula} via a hard-core phase shift in the P-wave, as
	\begin{align}
	\begin{aligned}
	\label{eqn:ps2}
	\delta_{rc}(x = q a_S) &=  \arctan \left(\frac{j_1(x)}{n_1(x)}\right) \\
						    &= \arctan \left(\frac{x -\tan{x}}{1+x \tan{x}}\right),
	\end{aligned}
	\end{align}

	\noindent and fix $a_S = 2 r_0 = 0.6 \, {\rm fm}$,  in order to compare with the excluded volume results. Here $a_S$ is identified as the closest distance of approach between two hard spheres. In this {\it scheme}, the thermodynamic pressure can be readily obtained from Eqs.~\eqref{eqn:p_bu}~-~\eqref{eqn:eff_sf}.

	As seen in Fig.~\ref{fig:one}, both approaches give comparable corrections to the thermodynamics. The effect is, as expected, much smaller due to the P-wave nature of the repulsion. However, in view of the empirical phase shift, even such a small modification is physically inconsistent. In Fig.~\ref{fig:two}, it is shown that the introduction of an additional repulsive force distorts the phase shift in the $P_{33}$ channel. This is shown together with the effective spectral function introduced in Eq.~\eqref{eqn:eff_sf}. From   Fig.~\ref{fig:two}, it is clear that the modification by the repulsive core is already substantial in the region of large center-of-mass energy.
%
%
	This, however, is unphysical, since the empirical phase shift should already contain all the attractive and repulsive interactions, regardless of their origin. Therefore, as far as the phase shift is concerned, the application of the additional excluded volume effect in the $P_{33}$ channel is redundant.

	\section{Conclusion}

	We have compared two approaches in modeling repulsive interactions. The excluded volume approach imposes the idea of closest-distance-of-approach among hadrons in a classical-statistical framework, while the S-matrix approach implements the same idea based on the quantum-mechanical problem of a hard-core potential.
	
	
	The major observation is that the introduction of an excluded volume between pions and nucleons, in addition to the interaction that generates the $\Delta$-resonance, leads to a distortion of the phase shift in the $P_{33}$ channel. This is unphysical since the empirical phase shift should already contains all the attractive and repulsive interactions, regardless of their origin.
	{
	Therefore, as far as the phase shift is concerned, there is no need to introduce an additional excluded volume in the $P_{33}$ channel.
	
	This suggests that the intricate repulsive force between hadrons is heavily interaction-channel dependent, and is hence unlikely to be captured by a single phenomenological parameter. Nevertheless, from a modeling perspective, the implementation of repulsive forces via the S-matrix formalism has the flexibility to accommodate a channel-dependent scattering length. This idea is supported by phenomenology , e.g. in the study of $\pi \pi$ and $\pi K$ scattering phase shifts in Ref. \cite{Ishida}, a strong isospin dependence of the hard-core radius parameter is indicated by the experimental data. The isospin $I = 2$ ($I = 3/2$) radius is found to be significantly smaller than the isospin $I = 0$ ($I = 1/2$) counterpart for the case of S-wave $\pi \pi$ ($\pi K$) scattering. This represents a further generalization of the approach where one associates an eigenvolume to a particle species \cite{Kapusta}, since in the current case the same particle could have a different effective hard-core radius, depending on the interaction channel.}

\acknowledgments

We acknowledge  stimulating discussions with P. Braun-Munzinger. B.F., P.M.L. and K.R. also  acknowledge fruitful  discussion with  R. Venugopalan and A. Andronic.
This work was partly supported by the Polish National Science Center (NCN), under Maestro grant DEC-2013/10/A/ST2/00106 and by the Extreme Matter Institute EMMI, GSI.

\end{document}